\documentclass[prb,aps,twocolumn,dcolumn,superscriptaddress,showpacs,amsfonts,amsmath,amssymb,showkeys,floatfix]{revtex4}
\usepackage{bm}
\usepackage{graphicx}% Include figure files
\begin{document}
\title{Theoretical simulation of the anisotropic phases of antiferromagnetic thin films}
\date{\today}
\author{Juan J. Alonso}
\affiliation{F\'{\i}sica Aplicada I, Universidad de M\'alaga,
29071-M\'alaga, Spain}
\email[E-mail address: ] {jjalonso@Darnitsa.Cie.Uma.Es}
\author{Julio F. Fern\'andez}
\affiliation{ICMA, CSIC and Universidad de Zaragoza, 50009-Zaragoza, Spain}
%\altaffiliation{IVIC}
\email[E-mail address: ] {JFF@Pipe.Unizar.Es}
\homepage[ URL: ] {http://Pipe.Unizar.Es/~jff}

\date{\today}
\pacs{75.10.-b, 75.30.Kz,75.70.Ak}
%\keywords{spin reorientation, films, anisotropy, Monte Carlo}

\begin{abstract}
We simulate antiferromagnetic thin films. Dipole-dipole and antiferromagnetic exchange interactions as well as uniaxial and quadrupolar anisotropies are taken into account. Various phases unfold as the corresponding parameters, $J$, $D$ and $C$, as well as the temperature $T$ and the number $n$ of film layers vary. We find (1) how the strength $\Delta_\nu$ of the anisotropy arising from dipole-dipole interactions varies with the number of layers $\nu$ away from the film's surface, with $J$ and with $n$; (2) a unified phase diagram for all $n$-layer films and bulk systems; 
(3) a layer dependent spin reorientation (SR) phase in which spins rotate continuously as $T$, $D$, $C$ and $n$ vary; (4) that the ratio of the SR to the ordering temperature depends (approximately) on $n$ only through $(D+\Delta /n)/C$, and hardly on $J$; (5) a phase transformation between two different magnetic orderings, in which spin orientations may or may not change, for some values of $J$, by varying $n$. 
\end{abstract}

\maketitle

\section{Introduction}
Magnetic thin films are attracting much interest. Some of it derives from applications (in electronics\cite{tappl}) of ferro/antiferromagnetic layered structures, where bias hysteresis arises from interactions at the interfaces of film\cite{ebh} and in nanoparticle layers.\cite{nano} Knowledge of the nature of the magnetically ordered states, as well as of the transitions between them, is important. The spin reorientation (SR) transition is most interesting. Continuous SR transitions, in which the direction of the magnetization changes continuously with temperature, were first observed in bulk ferrimagnets\cite{gill}
and in canted spin antiferromagnets (AFs).\cite{gyor}
Discontinuous SR transitions were first discovered in the bulk, in AFs\cite{neel} and in ferromagnets.\cite{morin} 
 
Competition of various magnetic anisotropies play decisive role in SR. Simply put, minimization of the energy
with respect to direction of the magnetization $m$ (or some staggered magnetization $m_s$ for AFs) gives the physical direction of $m$ at very low temperatures. By proper choice of the anisotropy, the energy minimum can be controlled, and thus the direction of $m$. Furthermore, in a Ginzburg-Landau like theory, the anisotropy constants can be made to vary with temperature, and thus the direction of $m$. This approach was first use by Horner and Varma\cite{var,LL} for continuous SR. Mean field, as well as MC calculations, also give continuous SR in the bulk.\cite{ours} {\it Discontinuous} SR, on the other hand, occur when one local minimum in the free energy, for some spin direction, suddenly (as, for instance, the temperature varies) becomes the global minimum, at the expense of another local minimum, for another spin direction.   

For films, thermally driven SR transitions, whose nature (continuous or discontinuous) was not clearly established, were first reported for ferromagnets by Pappas et al.\cite{pappas} Usually,\cite{allens} but not always,\cite{ni} SR proceeds from out of plane to in plane as the temperature increases. Variation of the number $n$ of layers can also lead to SR transitions. \cite{allens,ni,FO15} Manifestly smooth SR transitions have been recently observed in ferromagnetic thin films.\cite{garr, sell}
There are two main sources for the out of plane in-plane anisotropies in films: 
(1) missing bonds at surfaces can give rise to large local magneto-crystalline anisotropies then; (2) dipole-dipole interactions induce important anisotropies in magnetic films. In ferromagnets, dipolar fields drive spins to lie in plane, rather than out of plane, because dipolar field energies ($\sim m^2$) that obtain when $m$ is out of plane are thus avoided.\cite{gyorgy} Dipolar fields lead to stripe like domains  in thin films when magneto crystalline anisotropies favor spins to be out of plane.\cite{gyorgy,villa,benne} Growth of such stripes of in plane spins, at the expense of out of plane domains (or the other way around), as the temperature varies, leads to continuous SR in ferromagnetic films.\cite{vedme} No {\it continuous} SR transition is obtained if a homogeneous magnetization as well as only a lowest order uniaxial anisotropy are assumed,\cite{pheno1,pheno2} as has sometimes been done in mean field theory,\cite{usadel1,usadelx} MC\cite{usadelx,santa} and a renormalization group calculation.\cite{renor}

The behavior of antiferromagnetic films is qualitatively different, mainly because anisotropic effects that arise from dipolar fields in AFs are more subtle than in ferromagnets. In AFs, fields decay exponentially beyond the system's boundaries, as expected from the following simple argument. Consider an AF filling all space where $z<0$.
In the vacuum (i.e., where $z>0$), the magnetic field
${\bf h}({\bf r})$ follows from ${\bf h}({\bf r})=\nabla \phi ({\bf r})$, where $\phi ({\bf r})$ is a suitably defined field. Since $\phi ({\bf r})$ obeys Laplace's equation for $z>0$, $\phi ({\bf r})$ can be expanded therein, in obvious notation, as $\sum_{{\bf k}}a_{{\bf k}}\cos ({\bf k\cdot r}_\parallel)\exp (-\mid k\mid z)$. This much follows as well for ferromagnets. The difference between AFs and ferromagnets arises from the fact that whereas $a_{{\bf k}}\simeq 0$ for $ \mid k \mid <\mid {\bf G}\mid$ where ${\bf G}$ gives the periodicity of $\phi (x)$ near the surface of an AF, for ferromagnets, $\mid {\bf G}\mid$ scales with the inverse ferromagnetic domain size. In addition, the previous argument suggests that anisotropic effects from dipolar fields may also decrease exponentially, away from surfaces, {\it within} AFs. Important qualitative differences between anisotropies in ferro- and antiferromagnets arise from this. Unfortunately, relevant experiments,\cite{stohr,maat,hellwig} and MC work  for one-layer antiferromagnetic films have only recently been reported.\cite{macisaac1,abu} A discontinuous SR has been simulated in one layer films with a weak antiferromagnetic exchange\cite{macisaac1} (in which dipolar interactions are dominant) as well as with a strong one.\cite{abu} Because no high order (beyond quadratic) site anisotropy was taken into account, {\it continuous} SRs did not obtain. Finally, there is a mean field theory calculation for one-layer Heisenberg spin systems which include dipolar interactions as well as the lowest order uniaxial anisotropy\cite{deng} which also yields a thermally driven discontinuous SR. 

Our aim in this paper is to study (i) how the effective surface anisotropy that arises from dipolar interactions in magnetically ordered AF films varies with film thickness and with exchange strength, (ii) how the magnetic phases depend on film thickness, as well as on exchange, the uniaxial $D$ and quadrupolar $C$ anisotropy constants, (iii) how spins on surface layers behave with respect to spins on inner layers, and (iv) how the continuous and discontinuous SR temperatures depend on various parameters. 

The plan of the paper is as follows. The model is specified in Sec. \ref{model}. 
Section \ref{esa} is about antiferromagnetic ordering in the ground state and the unification that can be achieved between film and bulk phase behavior. This unification comes about because the anisotropy that arises from dipole-dipole interactions is, as surmised in the Introduction, a surface effect.   
In Sec. \ref{esahom} we define two general homogeneous spin configurations. By MC simulations, we show that all antiferromagnetically ordered phases, except for the SR phase,\cite{SRphase} conform to these configurations. One (the other one) 
general configuration holds for AF ordered states in which exchange (dipolar) interactions dominate. We derive the anisotropy energy in each of these two configurations coming from dipolar interactions. Monte Carlo results show that the resulting effective anisotropy decays exponentially fast with distance away from films surfaces.  In Sec. \ref{89}, the ground state continuous SR transition is studied. By MC simulations, we study how surface anisotropy arising from dipole-dipole interactions drive spin directions as a function of layer position. Section \ref{phases} is about thermal effects. In Sect. \ref{phaseshom} we report MC results for transitions between various homogeneous magnetic phases. One of them is the {\it discontinuous} SR transition. In addition, a transition between two ordered states, with the same spin alignment, is found as the number of film layers changes. In Sect. \ref{SR}, we study, by MC simulations, the thermally driven continuous SR transition. 
Defining an effective uniaxial anisotropy constant $D_{eff}$ that takes into account the dipole-dipole induced anisotropy, we show that the ratio of the SR transition temperature to the ordering temperature depends on $D_{eff}/C$, but depends hardly on the exchange constant, as long as it is antiferromagnetic. 

\begin{figure}[!ht]
\includegraphics* [width=80mm]{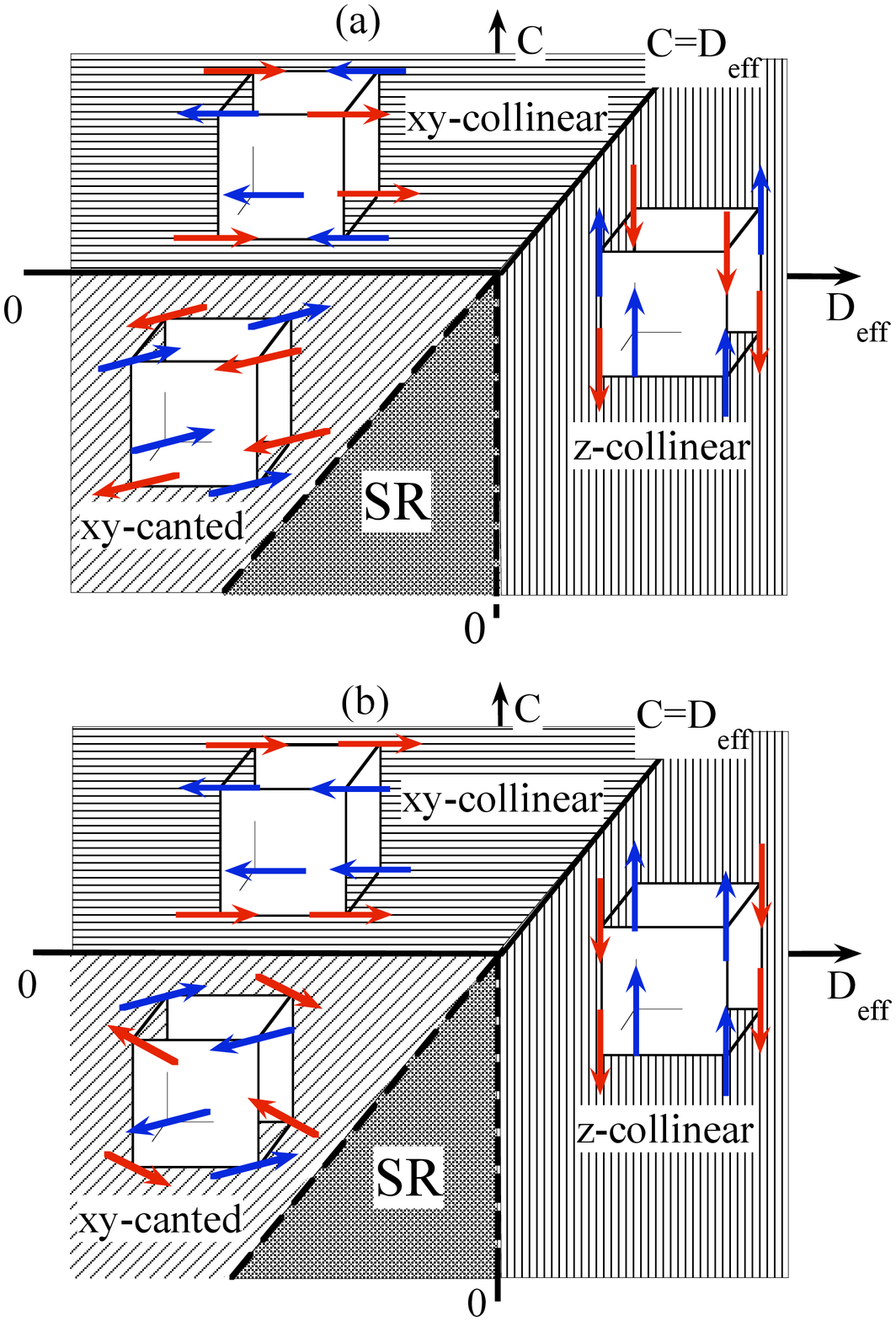}
\caption{(Color online) (a) Magnetic phase diagram for AFs in which exchange dominates.  In the SR phase, spins tilt away from the $z$-axis, $\phi =\pm\pi /4$, and all spins that are nearest neighbors to any one given spin point opposite to it.
(b) Same as in (a) but for AFs in which exchange is very weak and dipolar interactions dominate.
A spin configuration in the SR phase is depicted in Fig. 2. Full and dashed thick lines stand for first--
and second--order transitions, respectively. }
\label{f1}
\end{figure}

\section{The Model}
\label{model}
We next specify the model system we study. Let ${\bf S}_i$ be a 
classical 3-component unit spin at lattice site $i$ of a simple cubic (SC) lattice, let
\begin{equation}
{\cal H}={\cal H}_J+{\cal H}_d+{\cal H}_A,
\label{ham}
\end{equation}
where ${\cal H}_J=-J\sum_{\langle ij \rangle}{\bf S}_i\cdot {\bf S}_j$, the sum $\sum_{\langle ij \rangle}$ is over all nearest neighbor bonds,
\begin{equation}
{\cal H}_d=\sum_{\langle ij\rangle}\sum_{\alpha\beta}
T_{ij}^{\alpha\beta}S_i^\alpha S_j^\beta ,
\label{H_d}
\end{equation}

\begin{equation}
T_{ij}^{\alpha\beta}=\varepsilon_d
\left(\frac{a}{r_{ij}}\right)^3\left(\delta_{\alpha\beta}-3
\frac{r_{ij}^\alpha r_{ij}^\beta}{r_{ij}^2}\right),
\label{dipene}
\end{equation}
${\bf r}_{ij}$ is the displacement from site $i$ to site $j$, $a$ is the SC lattice parameter,
\begin{equation}
{\cal H}_A=-D\sum_i (S_i^z)^2-C\sum_i [(S_i^x)^4+(S_i^y)^4],
\label{A}
\end{equation}
and $D$ and $C$ are the uniaxial and quadrupolar anisotropy constants, respectively. The nearest neighbor dipolar energy $\varepsilon_d$ is defined through Eqs. (\ref{H_d}) and (\ref{dipene}).

The boundary conditions we use are most easily grasped in one dimension. Consider first spin sites at $x_k=ka$, for $k=-\infty\ldots , 0,\ldots \infty$. For periodic boundary conditions (PBC),
$S^{\alpha}_k=S^{\alpha}_{k+L}$ for all $k$, and we
let a spin at the $k$-th site interact with all $L/2$ ($L/2-1$) spins immediately to the right (left) of the $k$-th site. For free boundary conditions (FBC), on the other hand, we would let a spin at the $k$-th site interact with all spins on sites $n=1, \ldots , k-1, k+1,\ldots L$. We now return to the system of interest here,
an $n$-layer film, by which we mean $L\times L\times  n$ spins on a fully occupied SC lattice within a slab which lies flat on an $xy$ plane. Let the $z$ axis be perpendicular to the film layers.
We use PBC along the $x$ and $y$ directions and FBC along the $z$
direction.  Thus, a spin on any given site $i$ interacts, through dipolar fields, with all other $L\times L\times n -1$ spins in the system which are 
in a box, whose top and bottom surfaces coincide with the two films surfaces but is otherwise (that is, sidewise) centered on the $i$-th site. 

Our simulations follow the standard Metropolis Monte
Carlo  MC  algorithm.\cite{metrop} In particular, after we choose an initial 
spin configuration, we compute the
dipolar field at each site. Time evolution takes place as follows.
A spin is chosen at random and temporarily pointed in
a new random direction. The move is accepted if either $\Delta E\leq 0$, where
$\Delta E$ is 
the energy change, or with probability  $\exp (-\Delta E/k_BT)$,
where T is the systemÕs temperature, if $\Delta E>0$. All dipolar fields are then updated throughout the system if the move is accepted, before another spin is chosen to repeat the process. By {\it in-plane} and {\it out of plane} we will mean spins lying flat on the $xy$ plane or along the $z$ axis, respectively.

\section{Effective surface anisotropy from dipole-dipole interactions}
\label{esa}

The spin configurations explicitly depicted (not the SR phase) in Figs. 1(a) and 1(b) were shown in Ref. [\onlinecite{ours}] to be ground states for large $L\times L\times L$ systems with PBC. Our Monte Carlo calculations show that the same spin configurations are also ground states for films, with PBC at the film edges and FBC on the perpendicular direction to the film. In these states,
spins on the two film surfaces do not deviate at all from the direction they would point to in the bulk.\cite{farle} We shall refer to these states as {\it homogeneous}.
Our Monte Carlo calculations also show (see below) that, in the SR phase, spins on surface layers tilt away from these directions. We first derive the effective surface anisotropy that arises from dipole-dipole interactions in homogeneous states.

\subsection{Homogeneous states} 
\label{esahom}

Consider first the phases in Fig. \ref{f1}(a), in which,
\begin{equation}
{\textbf S}_i=(\sin \theta \cos \phi, \sin \theta \sin\phi, \cos \theta) \eta_i,
\label{cex}
\end{equation}
where
\begin{equation}
\eta_i\equiv (-1)^{x(i)+y(i)+z(i)},
\label{1astate}
\end{equation}
and $x(i),y(i),z(i)$ is the three dimensional position of the $i$-th site.
By proper choice of $\theta$ and $\phi$, the above equations define the three spin configurations shown in Fig. 1(a). We shall refer to these spin configurations, which minimize $E_J$, as $AF_J$ configurations. Spins in these configurations are clearly collinear.

Note first that Eqs. (\ref{cex}) and (\ref{1astate}) imply
\begin{equation}
E_J=(3-\frac{1}{n}) J,
\label{EJ0}
\end{equation}
independently of $\theta$ and $\phi$, and $E_A=-D\cos^2\theta -C (\sin^4\phi + \cos^4\phi )\sin^4 \theta$. To calculate $E_d$, we substitute Eqs. (\ref{cex}) and (\ref{1astate}) into Eq. (\ref{H_d}) and (\ref{dipene}).
Note first that $\sum_{ij}T_{ij}^{\alpha\beta}\eta_i\eta_j=0$ if $\alpha\neq \beta$ (by $x$ and $y$ reflection symmetry). Similarly, $\sum_{ij}T_{ij}^{xx}\eta_i\eta_j=\sum_{ij}T_{ij}^{yy}\eta_i\eta_j$. Therefore
\begin{equation}
E_d=(2N)^{-1}\sum_{ij}[T_{ij}^{xx}\eta_i\eta_j+(T_{ij}^{zz}-T_{ij}^{xx})\eta_i\eta_j\cos^2\theta ]
\label{Ed0}
\end{equation}
follows, which, by numerical evaluation, gives
\begin{equation}
E_d=\frac{0.67\varepsilon_d}{n}-\frac{\Delta}{n}\cos^2\theta
\label{De2} 
\end{equation}
where
\begin{equation}
\Delta =1.984\varepsilon_d
\label{E_j}
\end{equation}
for $AF_J$ spin configurations.

\begin{figure}[!ht]
\includegraphics*[width=80mm]{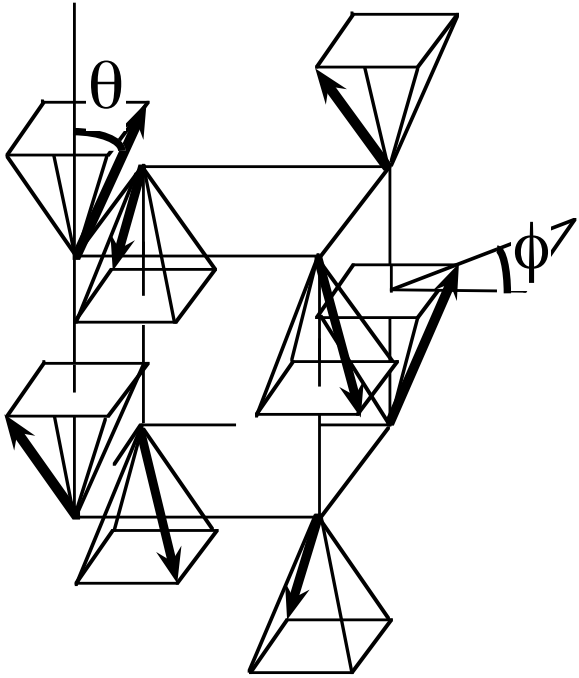}
\caption{Spin 
configuration
defined by Eqs. (\ref{ctx}) and (\ref{taud}). In the SR phase of Fig. 1(b),
$0<\theta <\pi /2$ and $\phi =\pm \pi /4$; for other phases of Fig. 1(b), $\theta$ and $\phi$ are as depicted therein.}
\label{canteado}
\end{figure}

We next calculate $E_J+E_d+E_A$ for the three spin configurations shown in Fig. 1(b), given by
\begin{equation}
{\textbf S}_i=(\tau_i^{x}\sin \theta \cos \phi, \tau_i^{y}\sin \theta \sin\phi, \tau_i^{z}\cos \theta)
\label{ctx}
\end{equation}
where $\bm{\tau}_i\equiv [\tau_i^x,\tau_i^y, \tau_i^z]$ is given by
\begin{equation}
\bm{\tau}_i=[(-1)^{y(i)+z(i)},
(-1)^{x(i)+z(i)},
(-1)^{x(i)+y(i)}].
\label{taud}
\end{equation}
We shall refer to the above spin configurations, depicted in Fig. 1b for some values of $\theta$ and $\phi$ and in Fig. \ref{canteado} for arbitrary $\theta$ and $\phi$, as $AF_d$ configurations. Spins in these configurations are in general noncollinear. It is worth pointing out that, in $L\times L\times L$ spin systems in cubic lattices, ${\cal H}_d$ is invariant with respect to both $\theta$ and $\phi$ in these $AF_d$ configurations.\cite{ours,Rinv}

Note first that Eqs. (\ref{ctx}) and (\ref{taud}) imply,
\begin{equation}
E_J=(1-\frac{1}{n}) J+\frac{2J}{n}\cos^2\theta,
\label{EJ}
\end{equation}
and $E_A=-D\cos^2\theta -C (\sin^4\phi + \cos^4\phi )\sin^4 \theta$.
We now calculate $E_d$.
For systems with complete cubic symmetry, that is, with a cubic lattice structure {\it and} a cubic shape, with the same type of boundary conditions on all surfaces, we have shown\cite{ours} that $E_d$ is invariant with respect to $\theta$ and $\phi$ in Eq. (\ref{ctx}). By the arguments preceding Eq. (\ref{De2}), we obtain 
\begin{equation}
E_d=(2N)^{-1}\sum_{ij}[T_{ij}^{xx}\tau_i^x\tau^x_j+(T_{ij}^{zz}\tau_i^z\tau_j^z-T_{ij}^{xx}\tau_i^x\tau^x_j)\cos^2\theta ]
\label{Ed1}
\end{equation}
for n-layer films with PBC at the edges and FBC on the top and bottom surfaces. Straightforward numerical calculations give 
\begin{equation}
E_d=-\left( 2.68-\frac{0.13}{n}\right)\varepsilon_d+\frac{1.23\varepsilon_d}{n}\cos^2\theta 
\label{Dep} 
\end{equation}
for $AF_d$ spin configurations. 

Thus, both for $AF_J$ and $AF_d$ configurations, an effective anisotropy,
\begin{equation}
D_{eff}=D+\Delta /n
\label{123}
\end{equation} 
obtains, where $\Delta$ is given by Eq. (\ref{E_j}) for $AF_J$, and by 
\begin{equation}
\Delta=-1.23\varepsilon_d-2J
\label{Delta_d}
\end{equation}
for $AF_d$ states. 

The anisotropy induced in AFs by dipole-dipole interactions, given by Eqs. (\ref{E_j}), (\ref{123}), and (\ref{Delta_d}),  differ from the one for ferromagnets in two respects: (1) it favors out of plane spins over in plane ones in $AF_J$ states, and (2) the corresponding energy for AFs varies as $1/n$ as $n$ increases. The reason for it is given in the Introduction, in plane orientation need not be favored in AFs, because no significant vacuum dipolar field energy exists for them. We next discuss the mechanism underlying the other effect, the $1/n$ behavior.

Equations (\ref{E_j}) and (\ref{Delta_d}) suggest that anisotropy effects arising from dipole-dipole interactions occur only on surface layers. We have (numerically) calculated how the dipolar field varies with the distance from a film's surface for homogeneous spin configurations. Let $\nu=0,1,2,\cdots$ number the layers, starting with $0$ for one of the two outermost surface layers. We find that the deviation of dipolar fields from their bulk value decreases exponentially as $\nu$ increases. More specifically,
\begin{equation}
\delta h_\nu\simeq \delta h\exp (-\kappa \nu),
\label{exp1}
\end{equation}
where $\kappa \simeq 4.4$ ($\kappa \simeq 7$) for all $AF_J$s and out of plane $AF_d$s (for all in plane $AF_d$'s) ordered states we have tried. This is in agreement with the discussion in Sec. I, since the wave vector's magnitude of the field near the surface of an $AF_J$ or an out of plane $AF_d$ (for an in plane $AF_d$) on a SC lattice is $\sqrt{2}\pi /a$ ($\sqrt{5}\pi /a$).
Therefore, to the accuracy of our numerical results, all anisotropy effects arising from dipole-dipole interactions occur only on the outer surfaces. Therein, it is given by $\Delta /2$ for each of the two surfaces on $n=2$ films, and by $\Delta$ for $n=1$ layer films. 

In order to characterize antiferromagnetically ordered states, we now define
\begin{equation}
m_{J}^\alpha =N^{-1}\sum_i  S_i^\alpha \eta_i,
\end{equation}
and 
\begin{equation}
m_{d}^\alpha =N^{-1}\sum_i  S_i^\alpha \tau^{\alpha}_i.
\end{equation}
In a $AF_J$ state, ${\textbf m}_{J}=(\sin \theta \cos \phi, \sin \theta \sin\phi, \cos \theta)$ and ${\textbf m}_{d}=0$. On the other hand,
in a $AF_d$ state ${\textbf m}_{J}=0$ and
${\textbf m}_{d}=(\sin \theta \cos \phi, \sin \theta \sin\phi, \cos \theta)$.
We can define these order parameters for the whole system, by summing over all sites $i$, or we can define, say, order parameters for surface film layers or interior layers, by summing over surface or interior sites.  
We next discuss the effective anisotropy in the SR phase.

\subsection{The SR phase}
\label{89}

\begin{figure}[!ht]
\includegraphics*[width=80mm]{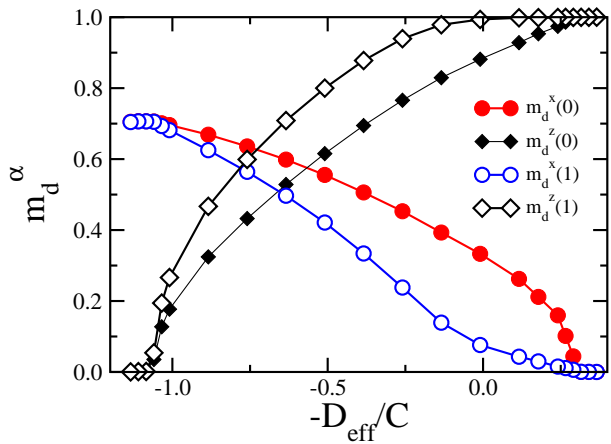}
\caption{(Color online) $m_d^\alpha $ versus $-D_{eff}/C$, for $\alpha =x, z$ on two different layers. $m_d^\alpha (\nu)$ is for the $\nu$-th layer ($\nu=0$ for a surface layer, and so on). All data follow from MC simulations of films of
$16 \times 16 \times n$ spins, for $n=4$, $C=-0.8\varepsilon_d$, and $J=0$, near $T=0$.}
\label{Fig3}
\end{figure}

\begin{figure}[!ht]
\includegraphics*[width=85mm]{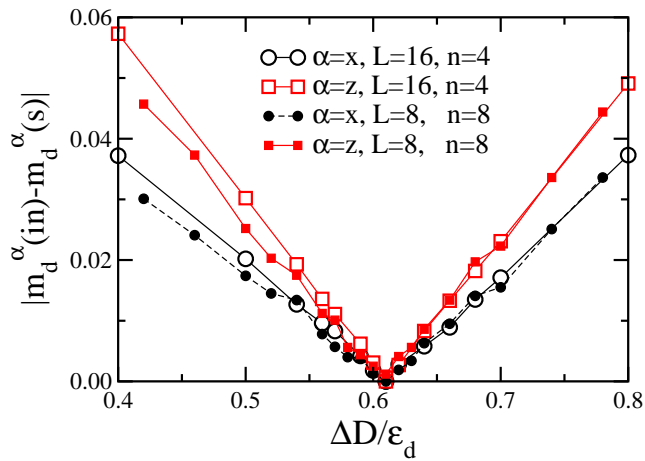}
\caption{(Color online) $|m_d^\alpha (in)-m_d^\alpha (s)|$ versus $\Delta D$ for
slabs of $L \times L \times n$ dipoles for $J=0$, $D=-\varepsilon_d$
and $C=-2 D$. We use FBC in the $z$ direction and PBC in the $xy$ plane.
The value of $D$ on the two outermost layers of the slab
differs by $\Delta D$ from the value it takes on the inner layers.}
\label{figure4}
\end{figure}

In Ref. [\onlinecite {ours}] the nature of the spin reorientation phase [marked as SR in Figs. 1(a) and 1(b)] in $L\times L\times L$ systems with PBC was discussed in some detail. In the SR phase $m_J^\alpha \neq 0$ ($m_d^\alpha \neq 0$) for all three components of $\alpha$ if $J\lesssim -1.34\varepsilon_d$ ($ -1.34\varepsilon_d\lesssim J\leq 0$), that is, spins in the SR phase
are tilted some angle $0<\theta <\pi /2$ away from the $z$ axis. For homogeneous states, given by either Eq. (\ref{cex}) or Eq. (\ref{taud}), as for $L\times L\times L$ systems in Ref. [\onlinecite{ours}], minimization of the total energy, gives
\begin{equation}
{\bf S}=(\pm u,\pm u, \pm v)
\label{Fdetb}
\end{equation}
for $C <D_{eff}<0$, where $ u=\sqrt{D_{eff}/2C}$ and $v=\sqrt{1-2u^2}$.
Then, varying $D_{eff}/C$ over the $0-1$ range would lead to a $\pi /2$ spin rotation. Inspection of Fig. \ref{Fig3} shows that the homogeneity assumption is wrong for thin films, and, consequently, spins do {\it not} quite rotate by $\pi /2$ as $D_{eff}/C$ sweeps over the $0-1$ interval.  
The order parameter, $m^x_d$ and $m^z_d$, on the surface clearly differs from the order parameters on inner layers in Fig. \ref{Fig3}. To the accuracy of our results, spins on all inner layers do follow either Eq. (\ref{cex}) or Eq. (\ref{ctx}). Thus, phase diagrams for n-layer films collapse into a single diagram if $D$ is replaced by $D+\Delta /n$, but only approximately so for the SR phase.

In order to look further into this effect, we have also performed MC simulations of n-layer films with anisotropy constants $D$ and $C$ on all sites, except for all surface sites, where a variable quantity $\Delta D$ is added to $D$. We calculate $\mid m_{d}^\alpha (in)\mid$ for inner layers, and $\mid m_{d}^\alpha (s)\mid$ for the two surface layers. A plot of 
$\mid m_{d}^\alpha (in) - m_{d}^\alpha (s)\mid$, for $\alpha =x, y,z$, vs $\Delta D/\varepsilon_d$ is shown in Fig. \ref{figure4} for $T\sim 0$,  $D=-\varepsilon_d$,
$C=-2 D$, and $J=0$. $\mid m_{d}^\alpha (in) -m_{d}^\alpha (s)\mid$ vanishes for all $\alpha$ at $\Delta D=0.615\varepsilon_d$, 
as was to be expected from Eqs. (\ref{Delta_d}) and (\ref{exp1}).
Further MC simulations we have performed for films of various thicknesses yield analogous results.

\section{phase transitions}
\label{phases}

Up to this point we have assumed which of the two, $AF_J$ or $AF_d$, states the system is in, but we are now able to specify which of these two obtain given the value of $J$. Much of this section, the portions having to do with phase transitions as the number of film layers change at very small temperatures, follow from the following considerations. 
Comparison of Eqs. (\ref{EJ0}), (\ref{De2}), and (\ref{E_j}) with (\ref{123}) and (\ref{Delta_d}) shows that: (1) $AF_J$ ($AF_d$) order ensues in out of plane spin configurations when $J\lesssim -1.34\varepsilon_d$ ($ -1.34\varepsilon_d\lesssim J <0$); $AF_J$ [$AF_d$] order ensues for in plane spin configurations when $J\lesssim -(1.34+0.27/n)\varepsilon_d$ [$ -(1.34+0.27/n)\varepsilon_d\lesssim J <0$]. From these conditions on $\mid J\mid$ one can decide whether $\mid J\mid$ is sufficiently small for a system to qualify as a dipolar antiferromagnet.\cite{dejongh}
As $n\rightarrow \infty$, the results obtained in Ref. [\onlinecite{ours}] for bulk systems follow. Note also that a transformation between $AF_J$ and $AF_d$ ordered states for in-plane configurations can occur as the number $n$ of layers changes if $-1.61\varepsilon_d\lesssim J\lesssim  -1.34\varepsilon_d$. 

\begin{figure}[!ht]
\includegraphics*[width=80mm]{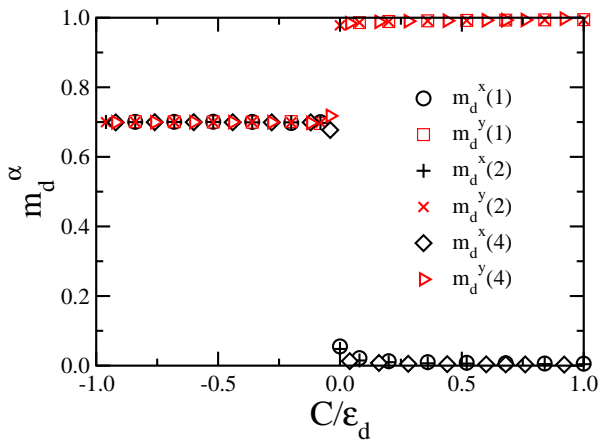}
\caption{(Color online) Order parameter $m_d^\alpha$, for
$\alpha =x, y$  vs $C$
for slabs of $16 \times 16\times n$ dipoles
on sc lattices at $T=0.05\varepsilon_d/k_B$,
where $k_B$ is Boltzmann's constant, for $D=-\varepsilon_d$ and $J=0$.
A transition between the $xy$-collinear and
$xy$-canted phases is clearly exhibited at $C=0$ for 
different values of $n$.
We use FBC in the $z$ direction and PBC in the $xy$ plane.
$C$ was lowered from $C=\varepsilon_d$ in 
$\Delta C=-0.02\varepsilon_d/k_B$ steps.
Each data point follows from averages over $10^5$ MC sweeps.}
\label{CantCol}
\end{figure}

\subsection{The C=0 transition}
\label{phaseshom}

\begin{figure}[!ht]
\includegraphics*[width=80mm]{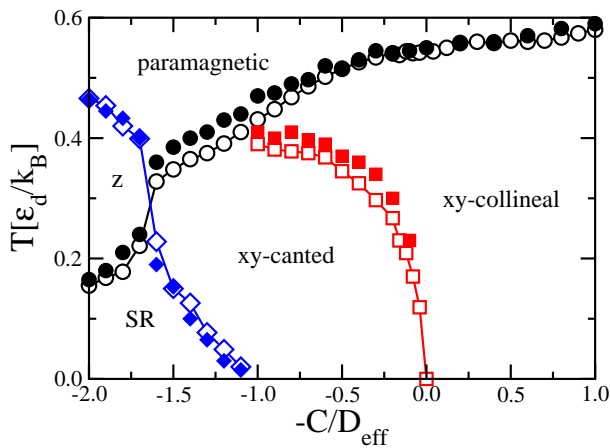}
\caption{(Color online) Transition temperatures vs $-C/D_{eff}$ from MC simulations of $L\times L\times 1$ spin systems, for $L=16$ and $L=32$, varying $C$, with $D_{eff}=-0.5\varepsilon_d$ and $J=0$ fixed. Open and closed symbols stand for $L=16$ and $L=32$, respectively.
$xy$ and $z$ stand for in plane and out of plane spin configurations. In the simulations,
the temperature was lowered in steps of $0.02\varepsilon_d/k_B$ and
$0.01 \varepsilon_d/k_B$ for $L=32$ and $L=16$, respectively. At each temperature, $10^6$ and $2\times 10^5$ MC sweeps were taken, for $L=16$ and $L=32$, respectively. In order to check for non-equilibrium effects, the system was {\it heated} for some values of $C/D_eff$. No such effects were found. }
\label{Tdep}
\end{figure}

The phase transition at $C=0$ for $D_{eff}<0$ is illustrated in Fig. \ref{CantCol}. We know of no previous experimental or MC work on this transition in films. 
It appears to be of first order, as predicted by
Landau's theory, because no symmetry group in any of these phases is a subgroup of another one. The transition moves slightly off the $C=0$ line as $T$ departs from 0, as shown for a one layer film in Fig. \ref{Tdep}, giving rise to a reentrant transition. The transition at $C=0$, however, remains unmoved at $T=0$ as the number of film layers varies. This is in agreement with the statement that dipolar interactions shift the value of $D$ but not of $C$.

\subsection{Discontinuous SR transitions}
\label{DSR}

\begin{figure}[!ht]
\includegraphics*[width=80mm]{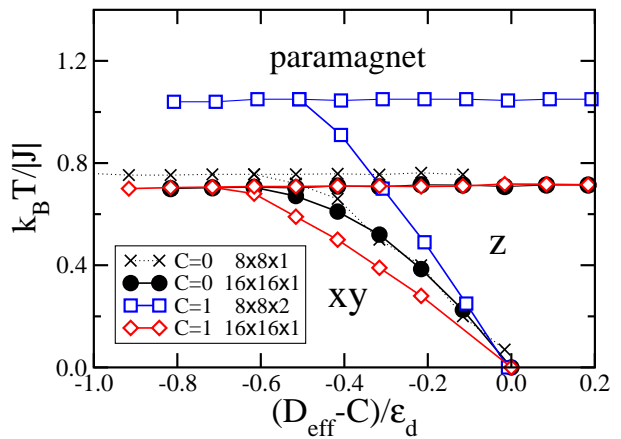}
\caption{(Color online) Transition temperatures vs $D_{eff}/C$ for $n=1$ and $n=2$ layer films in which $J=-10$, $C=0$ and $C=1$. In the graph, $z$ and $xy$ stand $z$-collinear and $xy$-collinear, respectively. All data points come from MC simulations of $L\times L\times n$ spins for the values of $L$ and $n$ shown in the graph. The phase transition boundaries at the top follow from the location of specific heat peaks obtained while lowering the temperature in $\Delta T=0.1$ steps.
In order to make sure equilibrium is realized, the lower phase transition boundary, for SR, is obtained from counting, over MC runs of several times $10^8$ sweeps, the frequency of occurrences of the two phases, which must be the same for both phases at the boundary line.}
\label{X}
\end{figure}

Consider first the phase transition between in plane and out of plane spin configurations at low temperature for $C>0$. Assume temporarily the same $AF_J$ or $AF_d$ order in both in plane and out of plane phases, and
recall that the ground state energy variation with spin tilt angle $\theta$ is given by ${\cal H}_A$ if $D$ is replaced by $D_{eff}$ in Eq. (\ref{A}). Then,\begin{equation}
{\cal H}_A=C\left[\left(\frac{D_{eff}}{C}-\sin^2\theta \right)\sin^2\theta\right] -D_{eff} ,
\label{A2}
\end{equation}
follows from Eq. (\ref{A}), replacing $D$ by $D_{eff}$ and assuming $S_x^2=1$ or $S_y^2=1$. Clearly, (1) $D_{eff}/C<1$ ($D_{eff}/C>1$) implies an in plane (out of plane) phase in the ground state, and (2) the transition is discontinuous since Eq. (\ref{A2}) gives an energy barrier between $\theta=0$ and $\theta=\pi /2$ when $D_{eff}=C$. This is  as depicted in Figs. 1(a) and 1(b). It has been observed in experimental and numerical work on films.\cite{allens,FO15} 
The assumption we made, that the same $AF_J$ or $AF_d$ order prevails in both in plane and out of plane phases, holds for most values of $J$, as follows from 
Eqs. (\ref{A}), (\ref{EJ0}), (\ref{De2}), (\ref{EJ}),  (\ref{Dep}), and (\ref{Delta_d}). More specifically, the transition occurs between two $AF_d$ states
if $-1,34\varepsilon_d\lesssim J\leq 0$. The transition is between two $AF_J$ states if $J\lesssim -(1.34+0.27/n)\varepsilon_d$.
Monte Carlo results that include this transition are shown in Fig. \ref{X} for films of $n=1$ and $n=2$.

The boundary line between out of plane and in plane phases tilts away from $D_{eff}=C$ as the temperature increases, as shown in Fig. \ref{X}.
Thus, the possibility of thermally driven SR transitions arise, as $T$ varies if $-0.5\varepsilon_d\lesssim D_{eff}-C< 0$. This is qualitatively as in the mean field prediction\cite{dos} for $J=-10^3\varepsilon_d$,
$C=0$, and $n=1$ in Ref. [\onlinecite{deng}].

 \begin{figure}[!ht]
\includegraphics*[width=80mm]{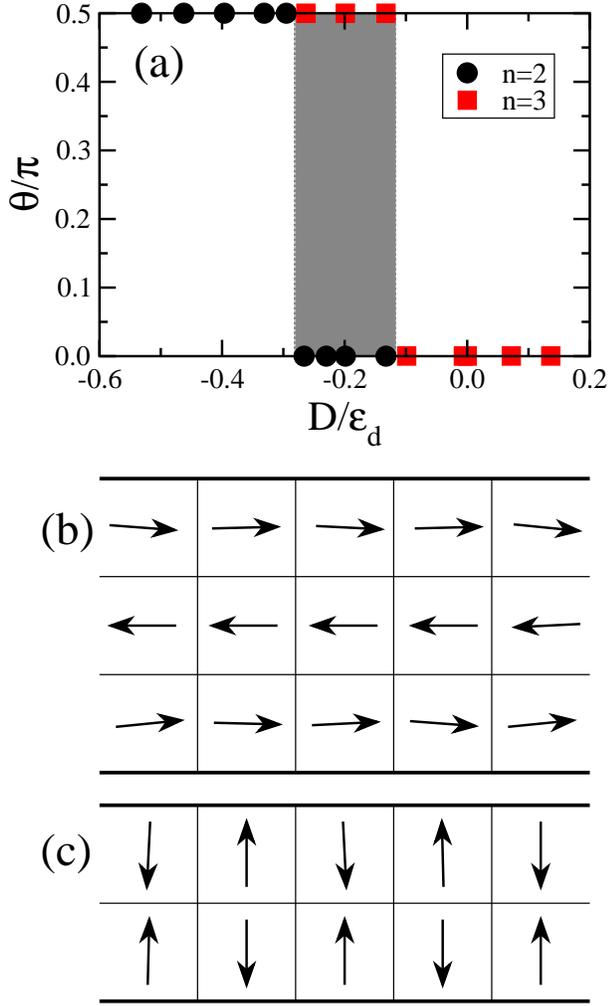}
\caption{(Color online) Everything shown is for $J=-1.38\varepsilon_d$, $C=0.5\varepsilon_d$, $k_BT=0.01\varepsilon_d$. This is the lowest temperature we obtained with MC simulations in which $T$ was lowered in  $\Delta k_BT=0.1\varepsilon_d$  steps, down to $k_BT=0.01\varepsilon_d$. Some $10^5$ MC sweeps were taken at each temperature step. (a) $\theta$ vs $D$ for $n=2$ ($\bullet$) and $n=3$ ($\blacksquare$). The shaded rectangle covers values of $D/\varepsilon_d$ where the out of plane $AF_J$ state, shown in (c) transforms into the the $AF_d$ in plane phase, shown in (b) if a layer is added to the film. 
(b) A $\theta =\pi /2$, $AF_d$, spin configuration, on lattice sites of a vertical cut of an $n=3$ layer film, obtained at $k_BT=0.1\varepsilon_d$ for $C=0.5$ and $D=-0.2\varepsilon_d$. (c) Same as in (b) but a $\theta =0$, $AF_J$, spin configuration that obtains for an $n=2$ layer film, for the same values of $C$, $D$, and $T$, as in (b). }
\label{ventana2}
\end{figure}

In the small $-(1.34+0.27/n)\varepsilon_d\lesssim J\lesssim -1.34\varepsilon_d$ range the situation is a more interesting. As specified at the top of Sect. \ref{phases}, the phase transition is then between an out of plane $AF_J$ ordered state and an in plane $AF_d$ ordered state. Equation  (\ref{A2}) does not apply then, because the assumption underlying it, that the same spin ordering $AF_J$ or $AF_d$ prevails on both sides of the phase boundary, breaks down. Assuming homogeneity, and making use of Eqs. (\ref{A}), (\ref{EJ0}), (\ref{De2}), (\ref{EJ}), and (\ref{Dep}), the condition for discontinuous SR transitions becomes
\begin{equation}
D=C+2J+(2.68-1.44/n)\varepsilon_d,
\label{notrep}
\end{equation}
for $C>0$. Note that quantity $D_{eff}$ is not well defined in this narrow $J$ range. We can however use the two different values $D_{eff}$ has on both sides of the phase transition for comparison of the energies of $AF_J$ and $AF_d$ ordered systems. It can be checked straightfordwardly that, again, there is an energy barrier between the $\theta =0$ and $\theta =\pi /2$ phases.
This transition, which has not, as far as we know, thus far been observed, is illustrated in Fig. \ref{ventana2} with MC results for $3$ and $2$ layer films. The dark rectangle shown in Fig. \ref{ventana2}, showing the range of values of $D/\varepsilon_d$ where $n=2$ ($n=3$) films order in out of plane $AF_J$ (in plane $AF_d$) states, follows from our MC simulations. It is slightly displaced to the left, by $D/\varepsilon_d\simeq 0.12$, from the prediction that follows from Eq. (\ref{notrep}). Irreversibility, that keeps spins from reorienting, from in plane to out of plane, at low $T$, as $T$ decreases, is responsible for this effect.

\subsection{Transitions between $AF_J$ and $AF_d$ ordered states}
\label{n}

\begin{figure}[!t]
\includegraphics*[width=85mm]{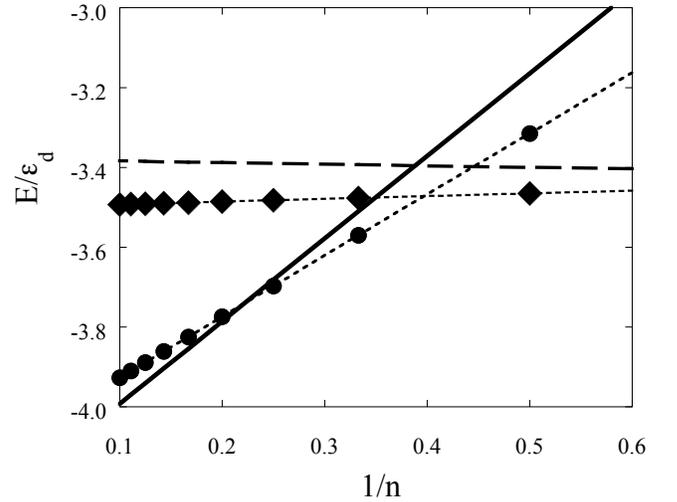}
\caption{Ground state energies vs $1/n$ for $AF_J$ and $AF_d$ states, for out-of and in-plane spin alignment, of $n$-layer slabs, with
$J=-1.38\varepsilon_d$, $D=-0.7\varepsilon_d$, and $C=0$.
$\bullet$ and full line stand for in-plane (spins either along the lattice axes or diagonally to them) $AF_d$ and $AF_J$ ordered states, respectively; dashed line, and $\blacklozenge$ stand for out of plane $AF_d$ and $AF_J$ ordered states, respectively. Dotted lines are guides to the eye.}
\label{Fig6}
\end{figure}

Phase transformations that do not involve SR can also occur between $AF_J$ and $AF_d$ ordered states, as the number $n$ of layers changes, if $-(1.34+0.27/n)\varepsilon_d\lesssim J\lesssim  -1.34\varepsilon_d$. 
This is illustrated in Fig. \ref{Fig6} for  
$J=-1.38\varepsilon_d$, $D=-0.7\varepsilon_d$, and $C=0$, where a transition from an $AF_J$ to an $AF_d$ ordered state, both in plane, is shown to take place as $n$ decreases from $n=6$ to $n=5$. (This is followed by a spin rotation, from in plane to out of plane, as $n$ decreases from $n=3$ to $n=2$.) We are not aware of any experimental observation of this kind of phase transformation.

\begin{figure}[!ht]
\includegraphics*[width=80mm]{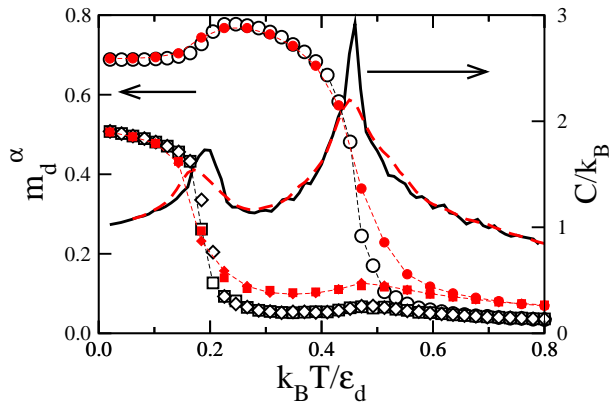}
\caption{(Color online)
$m_{d}^z$ ($\circ$ and $\bullet$),
$m_{d}^y$
($\square$ and $\blacksquare$),
$m_{d}^x$ ($\lozenge$ and $\blacklozenge$),
and $C/k_B$ (continuous and dashed lines) vs $T$.
Open (closed) symbols and continuous (dashed) lines are for 
MC results for systems of $32\times 32\times 1$ (of $16\times 16\times 1$)
spins on sc lattices for $J=0$,
$C=-\varepsilon_d$, and $D=0.7C$.
At each value of $T$, $4\times 10^5$ MC sweeps were
made. Lines are guides to the eye.}
\label{SRtr2}
\end{figure}

\subsection{Continuous SR transitions}
\label{SR}

A thermally driven SR transition is illustrated in Fig. \ref{SRtr2} for an $n=1$ layer film. It is rather similar to thermally driven transitions in $L\times L\times L$ systems with PBC.\cite{ours} However, as is pointed out in Sec. \ref{89} (and illustrated in Fig. \ref{Fig3}) for $T\sim 0$, the SR phase in films with $n>1$ is special. Spin configurations in the SR phase are not homogeneous. Whereas spins on inner layers follow Eqs. (\ref{cex}) or (\ref{ctx}), spins on surface layers do not, if $D$ and $C$ are homogeneous throughout the system. 

\begin{figure}[!ht]
\includegraphics*[width=85mm]{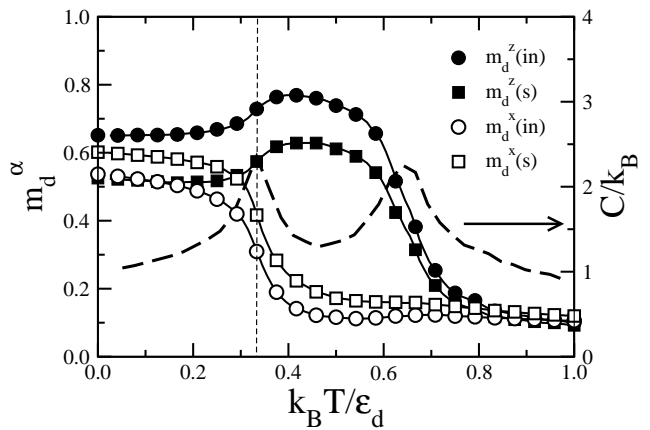}
\caption{Surface and inner layer order parameters $\mid m_{d}^\alpha (s)\mid$ and $\mid m_{d}^\alpha (in)\mid$, respectively, for $\alpha =x, y$, and $z$,
and specific heat, $C$, vs $T$. All data points come from MC simulations of of $8\times 8 \times 4$ spins with $J=0$. Symbols (dashed lines) are for order parameters (specific heat) of systems with spatially homogeneous anisotropy constants $D=-\varepsilon_d$ and $C=-2\varepsilon_d$ and FBC on the top and bottom surfaces; continuous lines are for films with PBC on all of its boundaries with a uniaxial anisotropy constant $D=-\varepsilon_d$ on inner surfaces, $D=-\varepsilon_d+\Delta /2$ on its top and bottom surfaces, and $C=-2\varepsilon_d$ everywhere.}
\label{pru}
\end{figure}

Inhomogeneity effects that arise from the effective surface anisotropy (induced by dipole-dipole interactions) are illustrated in Fig. \ref{pru} for a $4$-layer film as a function of temperature. Two kinds of data points from MC simulations are shown: (1) For films with spatially homogeneous anisotropy constants $D$ and $C$ and FBC on the top and bottom surfaces; (2) for films with PBC on all of its boundaries\cite{nota} with a uniaxial anisotropy constant that is $\Delta /2$ larger on its top and bottom surfaces than on the inner layers. Note how the order parameters on a surface layer differs from the order parameters on inner layers in the former case, and how direct application of an anisotropy $\Delta /2$ on surface layers of films with PBC (no anisotropy from dipolar interactions arises then) leads to the same effect.  This is as expected from the discussion in Sec. \ref{89}, concerning Fig. \ref{figure4}.

\begin{figure}[!ht]
\includegraphics*[width=80mm]{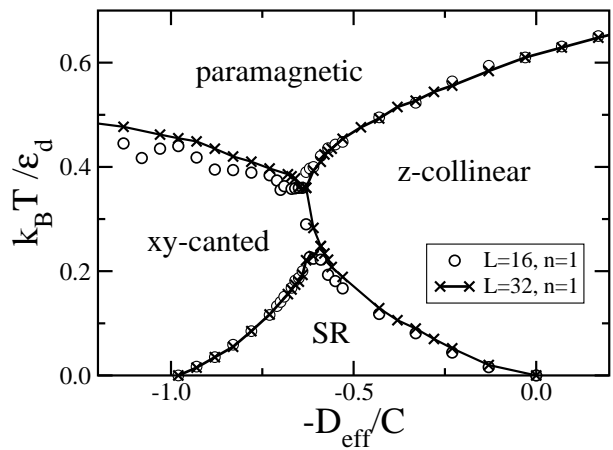}
\caption{(a) Phases of dipolar ($J=0$) antiferromagnetic
films for $C=-\varepsilon_d$. All
data points come from MC simulations of ($\circ$)
$16\times 16\times 1$ and ($\times$) $32\times 32\times 1$ spins.
$xy$-canted and $z$-collinear stand for the ordered phases that are depicted in Fig. 1(b); the SR phase is as pictured in Fig. 2.}
\label{SRph}
\end{figure}

The continuous SR portion of the phase diagram for films is rather like the one for bulk AFs.\cite{ours} We give the MC results we have obtained for $n=1$ layer dipolar films (in which $J=0$) in Fig. \ref{SRph}. 

\begin{figure}[!ht]
\includegraphics*[width=80mm]{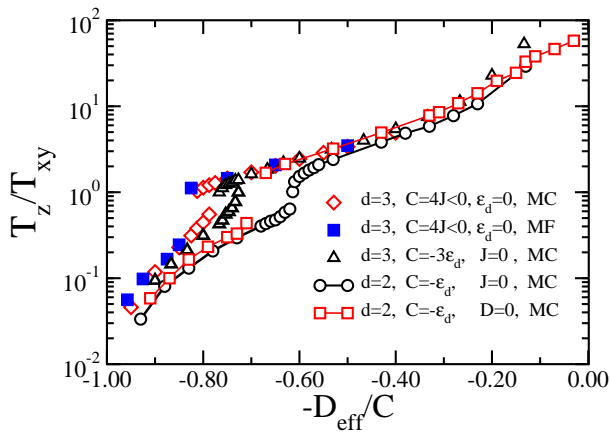}
\caption{(Color online) Temperature ratio $T_{z}/T_{xy}$
vs $-D_{eff}/C$ for the values shown of $\varepsilon_d$,
$J$, and $C$ in films ($d=2$) and $L\times L\times L$ systems 
with PBC ($d=3$). All data points, except the ($\blacksquare$) mean field (MF) results, come from MC simualtions. Note that $\circ$ and $\triangle$ stand
for pure dipolar systems ($J=0$), $\Diamond$ 
is for $\varepsilon_d=0$, and
$\square$ is for a one layer film in which both $D$ and $C$ are constant
($D_{eff}$ varies however, following Eqs. (\ref{123}) and (\ref{Delta_d}), because $J$ varies).
Systems are either of $L\times L\times L$ spins (for d=3) or of
$L\times L\times 1$ spins (for d=2). All symbols stand for $L=16$
except for $\circ$, which stands for $L=32$.
Lines are guides to the eye.}
\label{Tratio}
\end{figure}

Finally, we compare ordering and SR temperatures. Let $T_z$ ($T_{xy}$) stand for the temperatures below which $m_J^z\neq 0$ or $m_d^z\neq 0$  ($m_J^{xy}\neq 0$ or $m_d^{xy}\neq 0$). Monte Carlo results for the $T_z/T_{xy}$ ratio for $n=1$ films and bulk systems is shown in Fig. \ref{Tratio},
for $J/\varepsilon_d=0$ and for $\varepsilon_d/J=0$. In the latter case, $T_z/T_{xy}$ is, of course, independent of $J$, since $J$ sets the only energy scale then. The insensitivity of $T_z/T_{xy}$
to dimensionality and to the value of $J$, even when $\varepsilon_d\neq 0$ is remarkable.

\section{conclusions}

We have studied a nearest neighbor Heisenberg spin film like systems with dipolar interactions and uniaxial plus quadrupolar anisotropies. We have found how the strength $\Delta$ of the effective uniaxial anisotropy that arises from dipolar interactions varies with the strength of the antiferromagnetic exchange interaction and with layer position. We have argued (in Sec. I), and checked with MC simulations (in Sec. \ref{esahom}), that $\Delta$ decays exponentially fast with the distance from either of the two outermost film layers. We have also found (in Sec. \ref{esa}) that, except for the SR phase, all antiferromagnetic phases are homogeneous, that is, there are no surface states. These results, 
which are peculiar to AFs, imply that all but the SR phases of $n$ layer antiferromagnetic films fit into a single phase diagram [see Figs. 1(a) and 1(b)] if we let $D\rightarrow D+\Delta/n$. The only exception, the SR phase, associated with {\it continuous} SR (see Secs. \ref{89} and \ref{SR}), and phase transitions that arise from variation of film thickness, in the narrow $-1.61\varepsilon_d\lesssim J\lesssim -1.34\varepsilon_d$ range, in which spin orientation may (see Sec. \ref{DSR}) or may not (see Sec. \ref{n}) change. Finally, by means of MC simulations, we have found that the ratio of the continuous SR transition to the Ne\'el ordering temperature depends very little on $J$, as illustrated in Fig. \ref{Tratio}, and seems to depend on the number $n$ of film layers only through the effective uniaxial anisotropy constant $D+\Delta/n$.

\acknowledgments
{Financial support from Grant No. BFM2003-03919/FISI, 
from the Ministerio de Ciencia y 
Tecnolog\'{\i}a of Spain, is gratefully acknowledged.}


\begin{thebibliography}{99}
\bibitem{dos}Agreement is only qualitative, because whereas we see a SR transtion which near $T=0$ is at $D_{eff}\simeq C$, i.e. at $D=C-1.98\varepsilon_d$ for $J\ll \varepsilon_d$, from Eq. (\ref{E_j}), Deng et al., in Ref. [\onlinecite{deng}], find a value of $D$, for $C=0$, that is twice as large.
\end{thebibliography}
\end{document}